\documentclass[a4paper,11pt]{article}
\pdfoutput=1

\usepackage{jcappub}
\usepackage[T1]{fontenc}
\usepackage{amsmath,amssymb,bm}
\usepackage{graphicx}
\usepackage{booktabs}
\usepackage{xcolor}
\usepackage{float}
\graphicspath{{figures/}}

\newcommand{\LCDM}{\Lambda{\rm CDM}}
\newcommand{\dd}{\mathrm d}

\title{\boldmath Finding the distribution of matter using lenses -- II: deconvolution-based reconstruction with 3x2pt measurements}

\author[a]{Jun-Qian Jiang}

\author[b,c]{Ajoy Dawn}

\author[b,c,d]{Dhiraj Kumar Hazra}

\author[e]{Benjamin L'Huillier}

\author[a,f]{Arman Shafieloo}

\affiliation[a]{Korea Astronomy and Space Science Institute, Daejeon 34055, Korea}

\affiliation[b]{The Institute of Mathematical Sciences, HBNI, CIT Campus, Chennai 600113, India}
\affiliation[c]{Homi Bhabha National Institute, Training School Complex, Anushakti Nagar, Mumbai 400085, India}
\affiliation[d]{INAF/OAS Bologna, Osservatorio di Astrofisica e Scienza dello Spazio,
Area della ricerca CNR-INAF, via Gobetti 101, I-40129 Bologna, Italy
}

\affiliation[e]{Department of Physics and Astronomy, Sejong University, Seoul 05006, Korea}

\affiliation[f]{University of Science and Technology, Daejeon 34113, Korea}

\emailAdd{jiang@kasi.re.kr}
\emailAdd{ajoydawn@imsc.res.in}
\emailAdd{dhiraj@imsc.res.in}
\emailAdd{benjamin@sejong.ac.kr}
\emailAdd{shafieloo@kasi.re.kr}

\abstract{
We present a deconvolution-based framework for testing scale-dependent
departures of the late-time matter power spectrum from a fiducial cosmological
model using \(3\times2\)pt measurements.
We introduce a free-form scale-dependent modulation \(A(k)\) of the fiducial nonlinear matter power spectrum
and construct the
linear response of binned galaxy-clustering, galaxy--galaxy-lensing, and
cosmic-shear spectra to the discretized modulation \(A(k)\).  The response is
evaluated with full-sky, beyond-Limber kernels including density,
redshift-space-distortion, gravitational-shear, and intrinsic-alignment
contributions.  We reconstruct \(A(k)\) using a regularized modified
Richardson--Lucy algorithm, with weak diffusion in \(\ln k\) and selection of
the minimum-\(\chi^2\) solution along the iteration history.  Using Rubin/LSST
Year 10-like synthetic data, we find that oscillatory modulations with
amplitudes \(\gtrsim1\%\) can be recovered over
\(0.1\lesssim k\lesssim0.5\,{\rm Mpc}^{-1}\), provided the oscillation frequency
\(f\lesssim10\) on \(\log_{10}[k/(0.2\,{\rm Mpc}^{-1})]\).
We further introduce a posterior-weighted consistency statistic
calibrated with posterior-predictive null mocks, thereby accounting for
cosmological and nuisance-parameter uncertainties without relying on Wilks'
theorem.  The null case is consistent with \(A(k)=1\), while a \(1\%\) oscillatory modulation is
detected at \(\sim 2.6\sigma\).  These results demonstrate the potential of
regularized deconvolution as a model-independent consistency test of the matter
power spectrum in future \(3\times2\)pt surveys.
}

\begin{document}
\maketitle
\flushbottom

\section{Introduction}
\label{sec:intro}

The distribution of matter at late times is one of the most direct empirical
tests of the standard cosmological model.  In the minimal flat
\(\LCDM\) picture, the amplitude and shape of the matter power spectrum are
fixed by a small number of parameters that are tightly constrained by the
primary cosmic microwave background (CMB) anisotropies measured by
\textit{Planck}~\cite{Planck:2018vyg}.  The same model also gives an
economical description of background-distance measurements, including baryon
acoustic oscillations and type-Ia supernovae.
The growth of structure at lower redshift is a more
differential test: it probes the propagation of the CMB-inferred initial
conditions through nonlinear gravitational evolution, baryonic feedback,
galaxy bias, intrinsic alignments, and any possible departure from the standard
dark sector.

Weak-lensing measurements have played a central role in this test because they
probe the projected matter distribution with minimal assumptions about the
relation between galaxies and mass.  Stage-III cosmic-shear and
3x2pt analyses from DES Year~3~\cite{DES:2021wwk,DES:2021vln},
KiDS-1000~\cite{Heymans:2020gsg,Asgari:2020mwg}, and HSC
Year~3~\cite{Miyatake:2023njf} have generally preferred lower values of
\(S_8\equiv\sigma_8\sqrt{\Omega_{\rm m}/0.3}\) than the
\textit{Planck} \(\LCDM\) prediction.  Joint and later analyses have made this
picture more nuanced: DES Y3 and KiDS-1000 are mutually consistent in a common
cosmic-shear analysis~\cite{Kilo-DegreeSurvey:2023gfr}; KiDS-Legacy reports
cosmic-shear constraints closer to \textit{Planck} and passes extensive
internal consistency tests~\cite{Wright:2025xka,Stolzner:2025htz}; DES Year~6
cosmic shear still prefers a lower \(S_8\) than primary CMB constraints, but
with differences at the level of \(\lesssim2\sigma\) in the reported
comparisons~\cite{DES:2026mkc}.

Several recent works have promoted \(S_8\) comparisons to
direct or semi-direct inference of \(P_{\rm m}(k,z)\).  Amon and
Efstathiou~\cite{Amon:2022azi} introduced a phenomenological amplitude
\(A_{\rm mod}\) that interpolates between the linear and nonlinear matter power
spectra as a possible description of the weak-lensing suppression.  Preston
\textit{et al.}~\cite{Preston:2023uup,Preston:2024ggf} generalized this idea
to scale- and redshift-dependent suppression and studied the prospects for
future cosmic-shear surveys.  Broxterman and
Kuijken~\cite{Broxterman:2024oay} fit a double-power-law model in wavenumber
and scale factor to Stage-III weak-lensing data, finding that current surveys
mainly constrain an overall amplitude near a survey-dependent pivot.  A
regularized tomographic deprojection of KiDS-1000 cosmic shear reported
deviations from a purely cold-dark-matter reference spectrum, with the
interpretation depending on whether the reference amplitude is chosen near
\textit{Planck} or near the lower lensing-preferred \(S_8\)
\cite{Simon:2025xew}.  The corresponding KiDS-Legacy reconstruction finds
improved large-scale internal consistency with \(\LCDM\), while retaining a
small-scale suppression consistent with strong baryonic feedback
\cite{Broxterman:2025kpt}.  Ye, Jiang, and
Silvestri~\cite{Ye:2024rzp} proposed a model-independent expansion in powers of
the linear growth function, using multi-redshift data to separate linear and
nonlinear contributions where the data allow it.  More recently, CMB-lensing
and galaxy-lensing combinations have been used to reconstruct scale-dependent
deviations in the nonlinear matter spectrum: ACT DR6 CMB lensing combined with
DES Y3 cosmic shear indicates consistency with the CMB-inferred spectrum on
large scales while allowing mild scale-dependent suppression on smaller
scales~\cite{ACT:2023kun,Sarmiento:2025yyh}, and fast weak-lensing
reconstructions have begun to report, directly for DES, KiDS, HSC, and
ACT, posterior constraints on
\(\alpha(k)\equiv P(k)/P_{\rm fid}(k)-1\),
where \(P_{\rm fid}(k)\) denotes the nonlinear matter power spectrum
predicted by the reference cosmology adopted in the reconstruction
~\cite{Doux:2025vru}. These developments are also tied to the degeneracy
between baryonic feedback, non-standard dark matter, and modified growth:
matter-power-spectrum reconstruction has been advocated as a diagnostic for
distinguishing feedback from ultralight-axion suppression in future lensing
data~\cite{Preston:2025tyl}, while late-time growth-suppression models have
been studied as possible explanations of $S_8$ differences between HSC and CMB
\cite{Terasawa:2025fpf}.  In parallel, small-scale HSC analyses have tested
whether baryonic feedback alone can account for the lensing-inferred
suppression, finding no clear baryonic signature large enough to remove the
reported \(S_8\) offset~\cite{Terasawa:2024agq}.

The aim of this paper is complementary to these parameterized reconstructions.
We do not introduce a low-dimensional template for \(P_{\rm m}(k,z)\).  Instead,
we ask how much scale-dependent freedom in the matter power spectrum is
supported by the 3x2pt data.
We write the spectrum as a
modulation of a reference shape,
\begin{equation}
    P_{\rm m}(k,z_{\rm ref})
    =
    A(k)\,P_{{\rm m},0}(k,z_{\rm ref}),
\end{equation}
In this work, we use the reference redshift
\(z_{\rm ref}=0\)
and \(P_{{\rm m}}(k, z_{\rm ref})\) denotes the fiducial nonlinear matter power
spectrum of the unmodulated cosmology today.
The three components of the data vector, galaxy
clustering, galaxy--galaxy lensing, and cosmic shear, probe the same matter
field through different kernels, involving density, redshift-space distortion,
gravitational shear, and intrinsic-alignment contributions.  Their combination
therefore gives a linear projection of \(A(k)\).

To invert this projection we build on the Richardson--Lucy
algorithm~\cite{Richardson:1972hli,Lucy:1974yx}, originally developed for
image deconvolution.  In cosmology, Richardson--Lucy methods were adapted by
Shafieloo and Souradeep~\cite{Shafieloo:2003gf} to reconstruct the primordial
power spectrum from CMB anisotropies, and were subsequently developed for WMAP
and \textit{Planck} analyses of broad and oscillatory primordial features
\cite{Hazra:2013nca,Hazra:2014jwa}.  Sohn, Shafieloo, and
Hazra~\cite{Sohn:2022jsm} introduced regularized variants, including a
diffusion-based RegMRL update and a total-variation variant, to control the
noise amplification that is intrinsic to flexible deconvolution.  The first
paper in this series applied the same late-time deconvolution idea to CMB
lensing.  Here we apply the framework to
3x2pt measurements.

We demonstrate the framework on Rubin/LSST Year~10-like synthetic 3x2pt
data, using full-sky angular-power kernels and top-hat band powers.  The
full-sky calculation follows the SwiftC\(_\ell\) formalism~\cite{Reymond:2025ixl},
which evaluates beyond-Limber angular spectra efficiently with FFTLog-based
kernels while retaining scale-dependent growth.  We first show how RegMRL
responds to noisy injected oscillatory modulations of \(A(k)\).  We then
construct a posterior-weighted consistency statistic that marginalizes over the
cosmological and nuisance-parameter posterior and use it to test the
consistency of \(\LCDM\) with \(A(k)=1\).

The paper is organized as follows.  Section~\ref{sec:forward_model} introduces
the 3x2pt forward model, including the full-sky angular power spectra,
band-power binning, and the response matrix for \(A(k)\).
Section~\ref{sec:regmrl} describes
the RegMRL reconstruction and its behavior on injected oscillatory features.
Section~\ref{sec:consistency_tests} develops the posterior-weighted
consistency test and presents the null and weak-modulation calibrations.  We
summarize the results and limitations in Section~\ref{sec:conclusion}.
The survey configuration, priors, and fiducial synthetic-data
values are collected in Appendix~\ref{app:survey-config}.  Appendix~\ref{app:kgrid-frequency-checks}
examines the dependence of the reconstruction on modulation frequency using
additional logarithmic- and linear-phase injections.

\section{From the matter power spectrum to 3x2pt measurements}
\label{sec:forward_model}

We consider a 3x2pt data vector built from galaxy number counts and source
galaxy shapes.  For a lens tomographic bin \(i\), the number-count tracer is
written as a sum of density (\(D\)) and redshift-space-distortion
(\(R\)) contributions,
\begin{equation}
    g^i \rightarrow \{D^i,R^i\},
\end{equation}

while a source-shape F in bin \(i\) is written as
\begin{equation}
    \gamma^i \rightarrow \{G^i,I^i\},
\end{equation}
where \(G\) denotes gravitational shear and \(I\) denotes intrinsic
alignment.  The 3x2pt angular power spectra are then
\begin{subequations}
\begin{align}
    C_\ell^{gg,ij}
    &=
    C_\ell^{D D,ij}
    +C_\ell^{D R,ij}
    +C_\ell^{R D,ij}
    +C_\ell^{R R,ij},
    \label{eq:gg-split}
    \\
    C_\ell^{g\gamma,ij}
    &=
    C_\ell^{D G,ij}
    +C_\ell^{D I,ij}
    +C_\ell^{R G,ij}
    +C_\ell^{R I,ij},
    \label{eq:gs-split}
    \\
    C_\ell^{\gamma\gamma,ij}
    &=
    C_\ell^{G G,ij}
    +C_\ell^{G I,ij}
    +C_\ell^{I G,ij}
    +C_\ell^{I I,ij}.
    \label{eq:ss-split}
\end{align}
\end{subequations}
Magnification terms can be included in the same notation by adding a further
number-count contribution, but they are not part of the baseline 3x2pt model
used below.
CMB lensing can be incorporated with an analogous structure.  If the
auto- and cross-correlations among galaxy number counts, source shapes, and
CMB lensing are all included, the data vector becomes a 6x2pt combination.
The method developed here applies directly to that extended case, but in this
paper we focus on the 3x2pt subset.

\subsection{Full-sky contribution kernels}
\label{subsec:kernels}

Angular power spectra are often evaluated with the Limber approximation~\cite{Limber:1954zz,Kaiser:1991qi,Kaiser:1996tp,LoVerde:2008re}. 
The Limber approximation is computationally efficient and accurate for many
broad-kernel, high-multipole observables.  For future percent-level
measurements, however, large angular scales, narrow tomographic bins,
redshift-space distortions, and cross-correlations can make the Limber
approximation insufficient.  We therefore use a beyond-Limber calculation
following the SwiftC$_\ell$ prescription of Ref.~\cite{Reymond:2025ixl}, in
which the full-sky Bessel response is retained while the unequal-time matter
power spectrum is written in a factorized form.

For any pair of physical contributions \(a\) and \(b\), the full-sky angular
power spectrum can be written as
\begin{equation}
    C_\ell^{ab,ij}
    =
    \frac{2}{\pi}
    \int_0^\infty \dd k\,k^2
    \int \dd\chi\,\dd\chi'\,
    P_{\rm m}(k;\chi,\chi')\,
    \Delta_{\ell,a}^{i}(k,\chi)
    \Delta_{\ell,b}^{j}(k,\chi') .
    \label{eq:fullsky}
\end{equation}
Here \(\chi\) is comoving distance, \(P_{\rm m}(k;\chi,\chi')\) is the
unequal-time matter power spectrum, and \(\Delta_{\ell,a}^{i}\) is the
radial kernel and Bessel response for contribution \(a\).

The calculation uses a factorized unequal-time approximation,
\begin{equation}
    P_{\rm m}(k;\chi,\chi')
    =
    P_{\rm m}(k,z_{\rm ref})
    D(k,z(\chi))D(k,z(\chi')),
    \qquad
    D(k,z)
    =
    \left[
    \frac{P_{\rm m}(k,z)}
         {P_{\rm m}(k,z_{\rm ref})}
    \right]^{1/2},
    \label{eq:unequal-time}
\end{equation}
where $z_{\rm ref}=0$.
This turns Eq.~\eqref{eq:fullsky} into
\begin{equation}
    C_\ell^{ab,ij}
    =
    \frac{2}{\pi}
    \int_0^\infty \dd k\,k^2
    P_{\rm m}(k,z_{\rm ref})
    \Phi_{\ell,a}^{i}(k)
    \Phi_{\ell,b}^{j}(k),
    \label{eq:cell-factorized}
\end{equation}
with contribution-specific transforms
\begin{equation}
    \Phi_{\ell,a}^{i}(k)
    =
    \int \dd\chi\,
    D(k,z(\chi))\Delta_{\ell,a}^{i}(k,\chi).
    \label{eq:phi-def}
\end{equation}
The transforms entering the baseline 3x2pt calculation are
\begin{subequations}
\begin{align}
    \Phi_{\ell,D}^{i}(k)
    &=
    \int \dd\chi\,
    p_{\rm l}^{i}(\chi)\,
    b_i(\chi)\,
    D(k,z(\chi))\,
    j_\ell(k\chi),
    \label{eq:phi-density}
    \\
    \Phi_{\ell,R}^{i}(k)
    &=
    -
    \int \dd\chi\,
    p_{\rm l}^{i}(\chi)\,
    f(\chi)\,
    D(k,z(\chi))\,
    j_\ell''(k\chi),
    \label{eq:phi-rsd}
    \\
    \Phi_{\ell,G}^{i}(k)
    &=
    s_\ell
    \int \dd\chi\,
    W_G^{i}(\chi)\,
    D(k,z(\chi))\,
    \frac{j_\ell(k\chi)}{(k\chi)^2},
    \label{eq:phi-shear}
    \\
    \Phi_{\ell,I}^{i}(k)
    &=
    s_\ell
    \int \dd\chi\,
    p_{\rm s}^{i}(\chi)\,
    F_{\rm IA}(\chi)\,
    D(k,z(\chi))\,
    \frac{j_\ell(k\chi)}{(k\chi)^2},
    \label{eq:phi-ia}
\end{align}    
\end{subequations}
where
\begin{equation}
    s_\ell =
    \sqrt{\frac{(\ell+2)!}{(\ell-2)!}} .
\end{equation}
The normalized lens and source redshift distributions in comoving-distance
space are \(p_{\rm l}^{i}(\chi)\) and \(p_{\rm s}^{i}(\chi)\), \(b_i\) is the
linear galaxy bias, \(f=\dd\ln D_{\rm lin}/\dd\ln a\) is the growth rate used
in the RSD term, and \(F_{\rm IA}\) is the intrinsic-alignment transfer
amplitude.  In a spatially flat geometry the gravitational-shear kernel is
\begin{equation}
    W_G^{i}(\chi)
    =
    \frac{3H_0^2\Omega_{\rm m}}{2a(\chi)}
    \chi
    \int_{\chi}^{\chi_{\rm H}}
    \dd\chi_s\,
    p_{\rm s}^{i}(\chi_s)
    \frac{\chi_s-\chi}{\chi_s}.
    \label{eq:lensing-kernel}
\end{equation}
The factors of \((k\chi)^{-2}\) in Eqs.~\eqref{eq:phi-shear} and
\eqref{eq:phi-ia} represent the Poisson conversion between the potential-like
lensing response and the matter-density power spectrum.

\subsection{Band powers and the linear response to \texorpdfstring{\(A(k)\)}{A(k)}}
\label{subsec:linear-response}

The observed 3x2pt data vector consists of binned band powers.  For each
statistic \(X\in\{gg,g\gamma,\gamma\gamma\}\), tomographic pair \((i,j)\),
and band \(B\), we write
\begin{equation}
    \widehat C_{B}^{X,ij}
    =
    \sum_\ell
    \mathcal W_{B\ell}^{X,ij}
    C_\ell^{X,ij},
    \label{eq:bandpower}
\end{equation}
where \(\mathcal W_{B\ell}^{X,ij}\) is the bandpower window.
In this proof-of-concept study, we use TopHat windows.
After the
adopted scale cuts, the entries \(\widehat C_{B}^{X,ij}\) are stacked into a
single data vector \(\mathbf d\).  Here \(\mathbf d\) denotes the realized
observed or synthetic data vector, including noise.

In this work, we test a scale-only modulation of the nonlinear matter power
spectrum,
\begin{equation}
    P_{\rm m}(k,z)
    =
    A(k)P_{{\rm m},0}(k,z),
    \label{eq:aofk}
\end{equation}
where \(A(k)=1\) is the null model.
Here \(P_{{\rm m},0}(k,z)\) is the fiducial nonlinear matter power
spectrum predicted by the unmodulated model.
Under this ansatz the factor
\(D(k,z)\) in Eq.~\eqref{eq:unequal-time} is unchanged, while the reference
power spectrum in Eq.~\eqref{eq:cell-factorized} gains the multiplicative
factor \(A(k)\).  Discretizing \(A(k)\) on bins \(\mathcal K_\nu\), with
\(A(k)=a_\nu\) inside bin \(\nu\), the the deterministic theory band-power prediction is therefore a
linear function of \(\mathbf a\):
\begin{equation}
    \mathbf t(\vartheta,\mathbf a)
    =
    \mathbf G(\vartheta)\mathbf a,
    \qquad
    a_\nu \equiv A(k\in\mathcal K_\nu).
    \label{eq:linear-g}
\end{equation}
The parameter vector \(\vartheta\) denotes the cosmological and nuisance
parameters entering the kernels, the background geometry, and the fiducial
matter power spectrum.
The response matrix $\mathbf G$ has one row for each retained band power
\(m=(X,i,j,B)\) and one column for each \(k\)-bin \(\nu\).  Its explicit
entries are
\begin{subequations}
\begin{align}
    G_{(gg,i,j,B),\nu}
    &=
    \sum_\ell \mathcal W_{B\ell}^{gg,ij}
    \frac{2}{\pi}
    \int_{\mathcal K_\nu}\dd k\,k^2
    P_{{\rm m},0}(k,z_{\rm ref})
    \Big[
    \Phi_{\ell,D}^{i}(k)\Phi_{\ell,D}^{j}(k)
    +\Phi_{\ell,D}^{i}(k)\Phi_{\ell,R}^{j}(k)
    \notag\\
    &\hspace{4.2cm}
    +\Phi_{\ell,R}^{i}(k)\Phi_{\ell,D}^{j}(k)
    +\Phi_{\ell,R}^{i}(k)\Phi_{\ell,R}^{j}(k)
    \Big],
    \label{eq:g-matrix-gg}
    \\
    G_{(g\gamma,i,j,B),\nu}
    &=
    \sum_\ell \mathcal W_{B\ell}^{g\gamma,ij}
    \frac{2}{\pi}
    \int_{\mathcal K_\nu}\dd k\,k^2
    P_{{\rm m},0}(k,z_{\rm ref})
    \Big[
    \Phi_{\ell,D}^{i}(k)\Phi_{\ell,G}^{j}(k)
    +\Phi_{\ell,D}^{i}(k)\Phi_{\ell,I}^{j}(k)
    \notag\\
    &\hspace{4.2cm}
    +\Phi_{\ell,R}^{i}(k)\Phi_{\ell,G}^{j}(k)
    +\Phi_{\ell,R}^{i}(k)\Phi_{\ell,I}^{j}(k)
    \Big],
    \label{eq:g-matrix-gs}
    \\
    G_{(\gamma\gamma,i,j,B),\nu}
    &=
    \sum_\ell \mathcal W_{B\ell}^{\gamma\gamma,ij}
    \frac{2}{\pi}
    \int_{\mathcal K_\nu}\dd k\,k^2
    P_{{\rm m},0}(k,z_{\rm ref})
    \Big[
    \Phi_{\ell,G}^{i}(k)\Phi_{\ell,G}^{j}(k)
    +\Phi_{\ell,G}^{i}(k)\Phi_{\ell,I}^{j}(k)
    \notag\\
    &\hspace{4.2cm}
    +\Phi_{\ell,I}^{i}(k)\Phi_{\ell,G}^{j}(k)
    +\Phi_{\ell,I}^{i}(k)\Phi_{\ell,I}^{j}(k)
    \Big].
    \label{eq:g-matrix-ss}
\end{align}
\end{subequations}
The null prediction is
\begin{equation}
    \mathbf t_0(\vartheta)
    =
    \mathbf G(\vartheta)\mathbf 1.
\end{equation}

The survey definition, nuisance model, priors, and fiducial
synthetic-data parameters are collected in Appendix~\ref{app:survey-config}.

\section{Modified Richardson--Lucy reconstruction}
\label{sec:regmrl}

For a fixed \(\vartheta\), an observed vector \(\mathbf d\), response matrix
\(\mathbf G\), and covariance \(\mathbf C\), the iteration starts from
\(\mathbf a^{(0)}=\mathbf 1\).  At iteration \(n\),
\begin{align}
    \mathbf t^{(n)} &= \mathbf G\mathbf a^{(n)},\\
    \mathbf r^{(n)} &= \mathbf d-\mathbf t_\epsilon^{(n)},\\
    \mathbf q^{(n)} &= \mathbf C^{-1}\mathbf r^{(n)} ,
\end{align}
where \(m\) indexes retained band powers, \(t_{\epsilon,m}^{(n)}=
\max(t_m^{(n)},\epsilon)\), and \(\epsilon\) is the small positive floor used
for divisions and positivity constraints.  The response columns, indexed by
the \(k\)-bin label \(\nu\), are normalized by
\begin{equation}
    \widetilde G_{m\nu}
    =
    \frac{G_{m\nu}}{s_\nu},
    \qquad
    s_\nu =
    \begin{cases}
    \sum_m G_{m\nu}, & \left|\sum_mG_{m\nu}\right|>\epsilon,\\
    1, & \mathrm{otherwise}.
    \end{cases}
    \label{eq:column-normalization}
\end{equation}
Since the 3x2pt response matrix is not strictly positive, this is a
generalized Richardson--Lucy update rather than the classical positive-kernel
case.

The modified residual weights are
\begin{equation}
    W_m^{(n)}
    =
    \frac{r_m^{(n)}}{t_{\epsilon,m}^{(n)}}
    \tanh^2\!\left(q_m^{(n)}r_m^{(n)}\right),
    \label{eq:mrl-weight}
\end{equation}
and the multiplicative update for column \(\nu\) is
\begin{equation}
    u_\nu^{(n)}
    =
    \sum_m W_m^{(n)}\widetilde G_{m\nu}.
    \label{eq:mrl-raw-update}
\end{equation}
We use the bounded update
\begin{equation}
    \bar u_\nu^{(n)}
    =
    \mathcal C_{[-0.01,\,0.01]}\!\left(u_\nu^{(n)}\right),
    \qquad
    a_{{\rm base},\nu}^{(n+1)}
    =
    a_\nu^{(n)}\left(1+\bar u_\nu^{(n)}\right),
    \label{eq:mrl-clipped-update}
\end{equation}
where \(\mathcal C_{[a,b]}(x)=\min[\max(x,a),b]\) denotes clipping.
We set the clipping threshold for each update to 0.01, as we found that this value resulted in stable iterative evolution.
A regularization step is needed because this reconstruction is an
ill-conditioned deconvolution problem.  The angular band powers are broad
projections of the underlying matter power spectrum, so neighboring \(k\)-bins
have highly correlated response columns and some directions of \(\mathbf a\)
are only weakly constrained by the data.  Without a smoothness control, the raw
multiplicative update can therefore use noise fluctuations to generate rapidly
oscillating features in \(A(k)\).  We suppress these noise-dominated modes with
a weak diffusion term in \(\ln k\), while retaining the positive,
iteration-by-iteration structure of the Richardson--Lucy update.

The regularized version adds this diffusion term on the logarithmic wavenumber
grid \(x_\nu=\ln k_\nu\).  Let \(h_\nu=x_{\nu+1}-x_\nu\).  For interior bins
\(2\le\nu\le N_k-1\), the nonuniform-grid second-derivative operator is
\begin{equation}
    \left(\mathbf D_2\mathbf a\right)_\nu
    =
    \frac{2}{h_{\nu-1}+h_\nu}
    \left[
    \frac{a_{\nu+1}-a_\nu}{h_\nu}
    -
    \frac{a_\nu-a_{\nu-1}}{h_{\nu-1}}
    \right],
    \label{eq:d2-action}
\end{equation}
or, equivalently,
\begin{equation}
    (D_2)_{\nu,\nu-1}
    =
    \frac{2}{h_{\nu-1}(h_{\nu-1}+h_\nu)},
    \quad
    (D_2)_{\nu,\nu}
    =
    -\frac{2}{h_{\nu-1}h_\nu},
    \quad
    (D_2)_{\nu,\nu+1}
    =
    \frac{2}{h_\nu(h_{\nu-1}+h_\nu)} .
    \label{eq:d2-entries}
\end{equation}
All other entries in an interior row are zero, and the boundary rows are set to
zero so that the endpoints are not diffused.  RegMRL then updates
\begin{equation}
    \mathbf a^{(n+1)}
    =
    \max\left[
    \mathbf a_{\rm base}^{(n+1)}
    +\kappa \mathbf D_2 \mathbf a^{(n)},
    \epsilon
    \right],
    \label{eq:regmrl-update}
\end{equation}
where the maximum is applied component by component and \(\kappa\) controls the
strength of the smoothing.  Unless stated otherwise, the results use
\(N_{\rm iter}=100\) and \(\kappa=10^{-3}\).\footnote{
We checked the scale of this choice using the logarithmic \(k\)-grid and the
band-power resolution.  For the default grid,
\(\Delta\ln k=\ln(50/10^{-4})/159=0.0825\), so the explicit-diffusion part of
RegMRL has a stability scale
\(\kappa\lesssim(\Delta\ln k)^2/2\simeq3.4\times10^{-3}\).  The value used
here, \(\kappa=10^{-3}\), is below this scale and gives a characteristic
diffusion length \((2\kappa N_{\rm iter})^{1/2}\simeq0.45\) in \(\ln k\),
comparable to the effective resolution of the binned 3x2pt projection.}
We use $N_k-1=159$ bins uniformly spaced in $\ln k$ ($k \in [1 \times 10^{-4}, 50]$ Mpc$^{-1}$), ensuring that
the sampling rate exceeds the Nyquist requirement.
The reported reconstruction is
selected from the full iteration history by the best $\chi^2$ value,
\begin{equation}
    n_\star
    =
    \operatorname*{arg\,min}_{0\le n\le N_{\rm iter}}
    \left[
    \mathbf d-\mathbf G\mathbf a^{(n)}
    \right]^T
    \mathbf C^{-1}
    \left[
    \mathbf d-\mathbf G\mathbf a^{(n)}
    \right],
    \qquad
    \widehat{\mathbf a}_{\rm R}
    =
    \mathbf a^{(n_\star)} .
    \label{eq:best-reconstruction}
\end{equation}
The same best-\(\chi^2\) selection rule is applied to the data and to every
null mock.

The modulation used in our examples is an oscillatory
perturbation in \(\log k\), a feature shape widely studied in models and
searches for primordial spectrum oscillations
\cite{Chen2008,Flauger2010,Aich2013,ChenDvorkin2016,Braglia2021JCAP,
Braglia2022EPJC,Braglia2022PRD,Antony2026}, with a dimensionless amplitude
parameter \(\mathcal A_{\rm osc}\),
\begin{equation}
    A_{\rm osc}(k;\mathcal A_{\rm osc})
    =
    1
    +
    \mathcal A_{\rm osc}
    \sin\left[
    10\,\log_{10}\left(\frac{k}{k_\star}\right)
    \right],
    \qquad
    k_\star=0.2\,{\rm Mpc}^{-1}.
    \label{eq:oscillatory-modulation}
\end{equation}
The amplitude \(\mathcal A_{\rm osc}\) sets the fractional size of the
feature.

\begin{figure}[htp]
\centering
\includegraphics[width=0.82\textwidth]{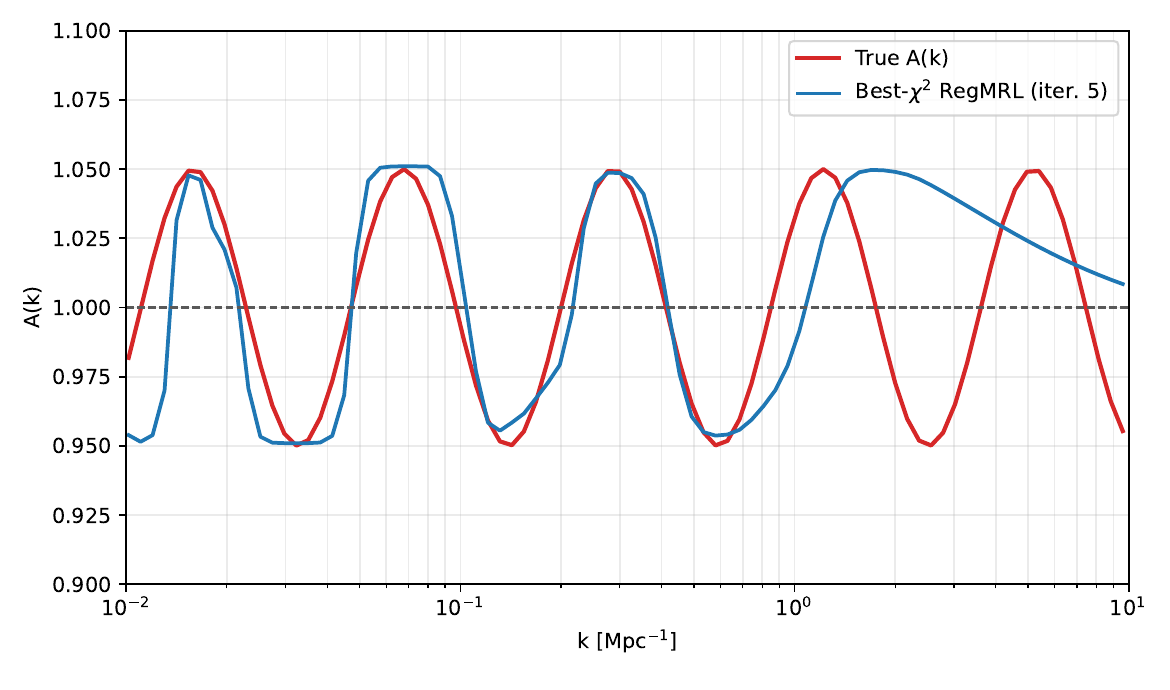}
\caption{Best-\(\chi^2\) RegMRL reconstruction for one noisy realization
of Eq.~\eqref{eq:oscillatory-modulation} with
\(\mathcal A_{\rm osc}=0.05\).  The reconstruction uses
\(N_{\rm iter}=100\), \(\kappa=10^{-3}\), and the update clipping in
Eq.~\eqref{eq:mrl-clipped-update}; the selected curve is the iteration with
the minimum $\chi^2$ along the trajectory.}
\label{fig:single-regmrl-bestchi2}
\end{figure}

Figure~\ref{fig:single-regmrl-bestchi2} shows one noisy realization generated
with \(\mathcal A_{\rm osc}=0.05\).  The RegMRL reconstruction tracks the
injected oscillatory structure in the \(k\)-range to which the 3x2pt kernels
are most responsive.

\begin{figure}[htp]
\centering
\includegraphics[width=0.82\textwidth]{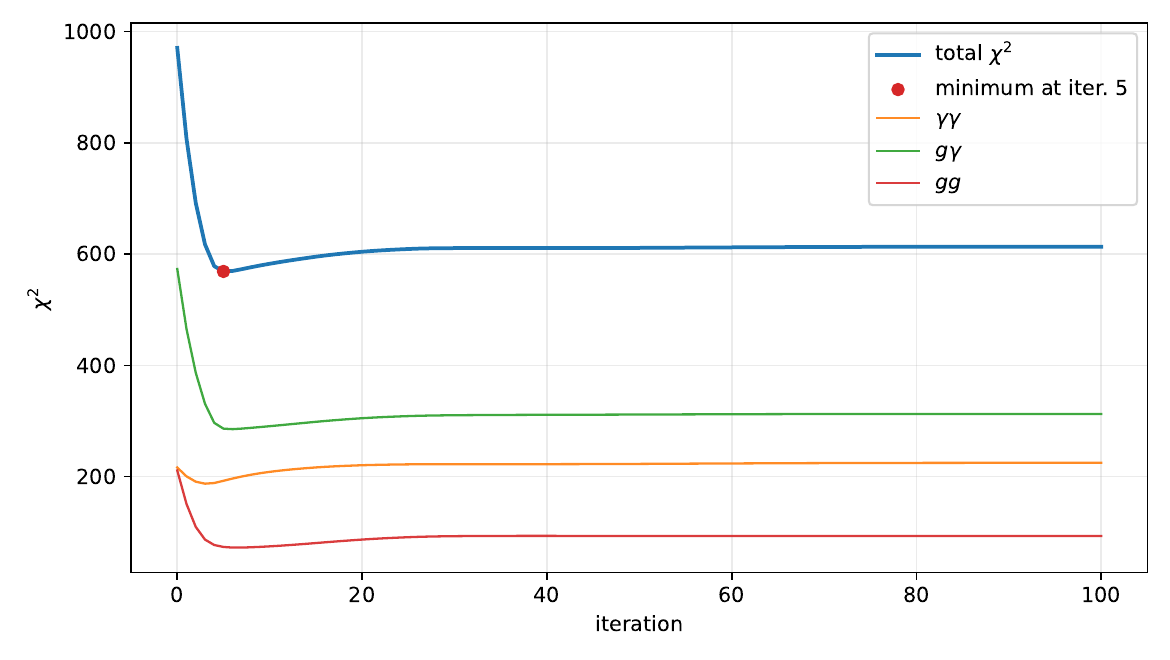}
\caption{Evolution of the total $\chi^2$ and its \(\gamma\gamma\),
\(g\gamma\), and \(gg\) contributions for the realization in
Fig.~\ref{fig:single-regmrl-bestchi2}.  The red marker indicates the
best-\(\chi^2\) iteration selected for the reported reconstruction.}
\label{fig:single-regmrl-chi2}
\end{figure}

The corresponding $\chi^2$ trajectory is shown in
Fig.~\ref{fig:single-regmrl-chi2}.  Most of the improvement occurs in the
first few iterations.  The best-\(\chi^2\) selection chooses iteration
\(n_\star=5\) for this realization, reducing the total $\chi^2$ from
\(\chi^2=970.23\) at \(A(k)=1\) to \(\chi^2=568.71\).
In the next section, we use this \(\Delta\chi^2\) as the starting point
for constructing our consistency-test statistic.
Continuing the
iteration to \(N_{\rm iter}=100\) gives \(\chi^2=613.10\).  After the
minimum, the total $\chi^2$ increases mildly as the multiplicative update
begins to follow directions that are less coherently supported by the projected
3x2pt data vector.  This behavior shows additional iterations do not necessarily represent additional
physical information.

\begin{figure}[htp]
\centering
\includegraphics[width=0.82\textwidth]{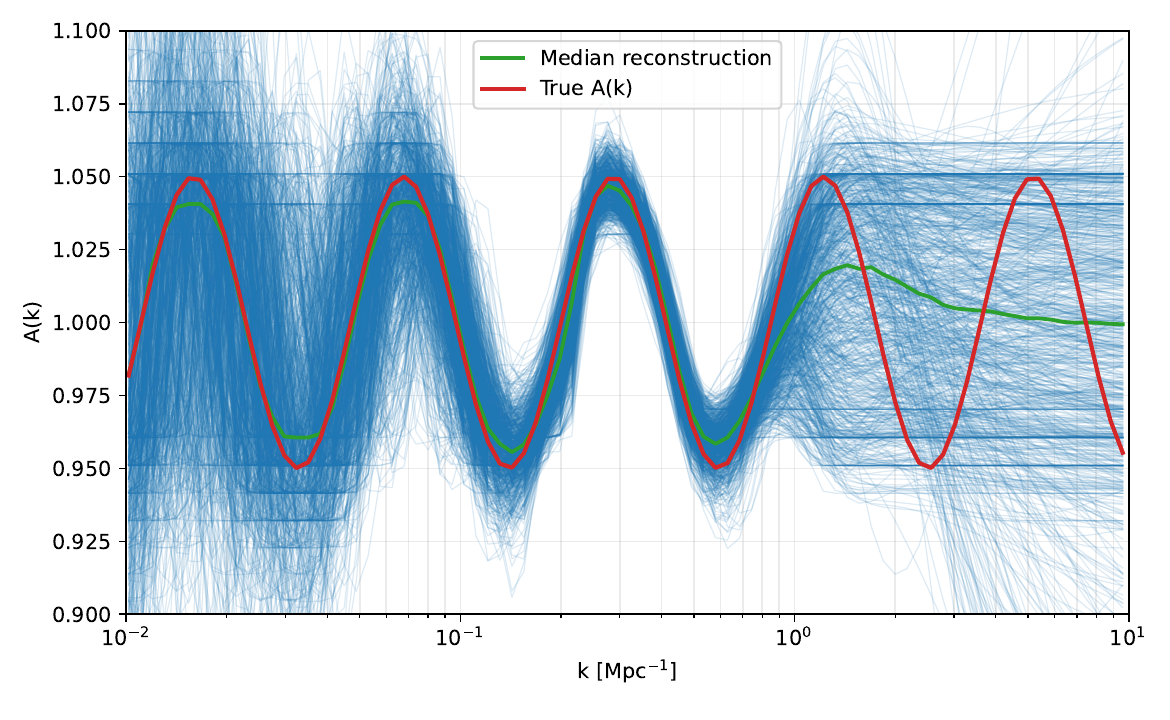}
\caption{RegMRL reconstructions for 1000 noisy realizations of
Eq.~\eqref{eq:oscillatory-modulation} with
\(\mathcal A_{\rm osc}=0.05\).  Each thin curve is the best-\(\chi^2\)
reconstruction for one realization; the thicker curve shows the median
reconstruction over the ensemble.  The ensemble shows reliable recovery over
\(10^{-1}\, {\rm Mpc}^{-1}\lesssim k\lesssim5\times10^{-1}\,{\rm Mpc}^{-1}\), while the
large scatter at smaller \(k\) limits the reliability of single-realization
reconstructions there.
}
\label{fig:regmrl-ensemble-fivepercent}
\end{figure}

\begin{figure}[htp]
\centering
\includegraphics[width=0.82\textwidth]{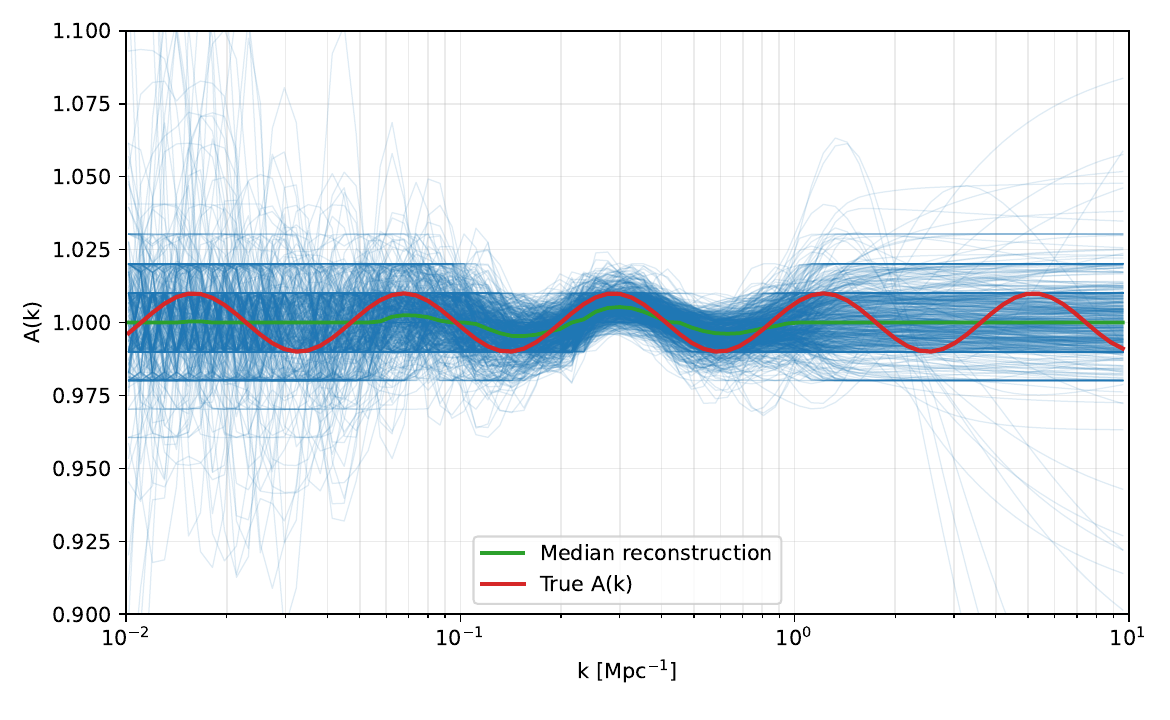}
\caption{Same as Fig.~\ref{fig:regmrl-ensemble-fivepercent}, but with
\(\mathcal A_{\rm osc}=0.01\).  This smaller-amplitude ensemble is used to
illustrate the weak-feature regime: a collective departure from \(A(k)=1\) is
visible over the best-constrained range, but many individual realizations
remain close to \(A(k)=1\).}
\label{fig:regmrl-ensemble-onepercent}
\end{figure}

Figures~\ref{fig:regmrl-ensemble-fivepercent} and
\ref{fig:regmrl-ensemble-onepercent} show reconstruction ensembles for
1000 noisy realizations, using the same RegMRL settings and selecting the
best-\(\chi^2\) iteration in each realization.  In the larger-amplitude case,
\(\mathcal A_{\rm osc}=0.05\), the noisy realizations indicate reliable
reconstruction over
\(1 \times 10^{-1}\, {\rm Mpc}^{-1}\lesssim k\lesssim5\times10^{-1}\,{\rm Mpc}^{-1}\):
Individual
curves track the oscillatory structure coherently through this range.  On
larger physical scales, \(k\lesssim10^{-1}\,{\rm Mpc}^{-1}\), the ensemble
median can still be compatible with the input \(A(k)\), but the large scatter
among noisy realizations shows that this part of the reconstruction is not
reliable for an actual data set, which would correspond to one realization
rather than to the ensemble median.  In the smaller-amplitude case,
\(\mathcal A_{\rm osc}=0.01\), the ensemble as a whole shows a coherent
departure from \(A(k)=1\) only over roughly the same
\(1 \times 10^{-1}\)--\(5\times10^{-1}\,{\rm Mpc}^{-1}\) interval, and the original oscillation
amplitude is not necessarily recovered.  Several noisy realizations already
remain close to \(A(k)=1\), indicating that a \(1\%\) modulation is near the
sensitivity limit of this observational configuration for the assumed
oscillatory shape of \(A(k)\).
Appendix~\ref{app:kgrid-frequency-checks}
further shows that this sensitivity requires the modulation frequency not to
be too high (frequency \(\lesssim 10\) on \(\log_{10}(k/0.2\,{\rm Mpc}^{-1})\)).

\section{Deconvolution reconstruction as a consistency test}
\label{sec:consistency_tests}

The visual recovery of an injected feature is not, by itself, a calibrated
consistency test.  First, fixing \(\vartheta\) ignores the
posterior uncertainty of the cosmological and nuisance parameters and hence of
the projection kernels in \(\mathbf G(\vartheta)\).  Second, interpreting the
reconstruction one \(k\)-bin at a time can both overstate the significance of
localized excursions through a look-elsewhere effect and understate the
significance of coherent, correlated shifts spread over many bins.  We
therefore use scalar improvement statistics, calibrated against null mocks, as
the consistency tests below.
The null hypothesis is
\begin{equation}
    \mathrm{Hyp}_0:\quad A(k)=1.
\end{equation}
For fixed \(\vartheta\), the null $\chi^2$ is
\begin{equation}
    \chi^2_0(\vartheta;\mathbf d)
    =
    \left[
    \mathbf d-\mathbf G(\vartheta)\mathbf 1
    \right]^T
    \mathbf C^{-1}
    \left[
    \mathbf d-\mathbf G(\vartheta)\mathbf 1
    \right].
    \label{eq:null-chi2}
\end{equation}
The conditional deconvolution improvement is
\begin{equation}
    T_{\rm best}(\vartheta;\mathbf d)
    =
    \chi^2_0(\vartheta;\mathbf d)
    -
    \chi^2_{\rm R,best}(\vartheta;\mathbf d),
    \label{eq:tbest}
\end{equation}
where \(\chi^2_{\rm R,best}\) is evaluated at
\(\widehat{\mathbf a}_{\rm R} = \mathbf a^{(n_\star)}\) (Eq. \ref{eq:best-reconstruction}).
We do not assign a Wilks-theorem
\(p\)-value to \(T_{\rm best}\).  The effective degrees of freedom depend on
the regularization, clipping, iteration count, and best-\(\chi^2\) selection
rule, so the statistic must be calibrated with null mocks analyzed in exactly
the same way.

The diagnostic averages the conditional improvement over the null
posterior,
\begin{equation}
    p(\vartheta|\mathbf d)
    \propto
    \mathcal L_0(\mathbf d|\vartheta)\pi(\vartheta).
\end{equation}
Here \(\mathcal L_0(\mathbf d|\vartheta)\propto
\exp[-\chi_0^2(\vartheta;\mathbf d)/2]\) is the null-model likelihood with
\(A(k)=1\), and \(\pi(\vartheta)\) is the joint prior on the cosmological and
nuisance parameters, given by Eqs.~\eqref{eq:survey-cosmo-priors} and
\eqref{eq:survey-nuisance-priors}.
The cosmological priors
are conservative and do not impose the Planck constraint.
The target statistic is the posterior-integrated likelihood-ratio improvement,
\begin{equation}
    S_{\rm best}(\mathbf d)
    =
    2\log\left[
    \int \dd\vartheta\,
    p(\vartheta|\mathbf d)
    \exp\left(
    \frac{T_{\rm best}(\vartheta;\mathbf d)}{2}
    \right)
    \right].
    \label{eq:sbest-continuous}
\end{equation}

In practice, we estimate Eq.~\eqref{eq:sbest-continuous} with MCMC samples.
We use an ordinary chain at temperature \(\mathcal T=1\) and a tempered
proposal at \(\mathcal T=2\),
\begin{equation}
    q_{\mathcal T}(\vartheta|\mathbf d)
    \propto
    \mathcal L_0(\mathbf d|\vartheta)^{1/\mathcal T}\pi(\vartheta).
\end{equation}
The data statistic is evaluated first on the \(\mathcal T=1\) chain.
If \(\vartheta_r^{(1)}\) are samples
from \(q_1(\vartheta|\mathbf d)\), with chain weights \(w_r^{(1)}\), then for
each sample we run the best-\(\chi^2\) RegMRL reconstruction on the observed
data vector and compute
\begin{equation}
    T_{{\rm best},r}^{\rm obs}
    =
    T_{\rm best}(\vartheta_r^{(1)};\mathbf d).
\end{equation}
The observed statistic is
\begin{equation}
    S_{\rm best}^{\rm obs}
    =
    2\left[
    \log\sum_r
    w_r^{(1)}
    \exp\left(\frac{T_{{\rm best},r}^{\rm obs}}{2}\right)
    -
    \log\left(\sum_r w_r^{(1)}\right)
    \right].
    \label{eq:sbest-obs}
\end{equation}

The same \(\mathcal T=1\) chain is then used to generate the
posterior-predictive mock hypotheses.  We draw \(N_{\rm mock}\)
post-burn-in samples from the \(\mathcal T=1\) chain,
\begin{equation}
    \vartheta_b^{\rm gen}\sim q_1(\vartheta|\mathbf d),
\end{equation}
and generate a noisy null data vector at each draw,
\begin{equation}
    \mathbf d_b
    =
    \mathbf G(\vartheta_b^{\rm gen})\mathbf 1
    +
    \bm\epsilon_b,
    \qquad
    \bm\epsilon_b\sim \mathcal N(\mathbf 0,\mathbf C).
    \label{eq:posterior-predictive-mock}
\end{equation}

For each mock, the \(\mathcal T=2\) chain is used only as an overdispersed
proposal.  We first cool the \(\mathcal T=2\) samples to the physical
posterior for the observed data, with weights proportional to
\begin{equation}
    w_s^{(d)}
    \propto
    w_s^{(\mathcal T)}
    \exp\left[
    \left(1-\frac{1}{\mathcal T}\right)
    \log \mathcal L_0(\mathbf d|\vartheta_s)
    \right],
    \qquad \mathcal T=2.
    \label{eq:cooling}
\end{equation}
and then reweight those samples to the mock posterior,
\begin{equation}
    w_{b,s}
    \propto
    w_s^{(d)}
    \exp\left[
    -\frac12
    \left(
    \chi^2_0(\vartheta_s;\mathbf d_b)
    -
    \chi^2_0(\vartheta_s;\mathbf d)
    \right)
    \right],
    \label{eq:mock-reweight}
\end{equation}
For each mock and each proposal sample \(\vartheta_s^{(2)}\), we run the same
best-\(\chi^2\) RegMRL reconstruction on \(\mathbf d_b\) and compute
\begin{equation}
    T_{{\rm best},b,s}
    =
    T_{\rm best}(\vartheta_s^{(2)};\mathbf d_b).
\end{equation}
The mock statistic is then
\begin{equation}
    S_{{\rm best},b}
    =
    2\left[
    \log\sum_s
    w_{b,s}
    \exp\left(\frac{T_{{\rm best},b,s}}{2}\right)
    -
    \log\left(\sum_s w_{b,s}\right)
    \right].
    \label{eq:sbest-mock}
\end{equation}
The posterior-predictive tail probability is
\begin{equation}
    p_{\rm PPC}
    =
    \frac{
    1+\sum_{b=1}^{N_{\rm mock}}
    \mathbf 1(S_{{\rm best},b}\ge S_{\rm best}^{\rm obs})
    }{
    1+N_{\rm mock}
    }.
    \label{eq:ppc-pvalue}
\end{equation}
Since large values of \(S_{\rm best}\) correspond to stronger deviations from
the null model, this is an upper-tail, one-sided probability.  When quoting an
equivalent Gaussian significance we therefore use
\begin{equation}
    Z_{\rm 1s}=\Phi^{-1}(1-p_{\rm PPC}),
    \label{eq:one-sided-sigma}
\end{equation}
where \(\Phi\) is the cumulative distribution function of a standard normal
random variable.
We also monitor the likelihood-ratio effective sample size.  For a mock \(b\), this
is
\begin{equation}
    N_{{\rm eff},b}
    =
    \frac{
    \left[\sum_s w_{b,s} \exp(T_{{\rm best},b,s}/2)\right]^2
    }{
    \sum_s \left[w_{b,s}\exp(T_{{\rm best},b,s}/2)\right]^2
    }.
    \label{eq:neff-lr}
\end{equation}
The posterior-weighted calculation is reported for two synthetic cases: a null
case generated with \(A(k)=1\), and a weak-modification case generated with
Eq.~\eqref{eq:oscillatory-modulation} using
\(\mathcal A_{\rm osc}=0.01\).  For both cases, \(S_{\rm best}^{\rm obs}\)
is computed on the corresponding \(\mathcal T=1\) chain using
Eq.~\eqref{eq:sbest-obs}.  The same \(\mathcal T=1\) chain supplies the
noisy mock hypotheses in Eq.~\eqref{eq:posterior-predictive-mock}.  We use
\(N_{\rm mock}=1000\) for the null case.
Because the \(p\)-value is smaller
for the 1\% modulation case, we increase the ensemble to
\(N_{\rm mock}=10000\) in this proof-of-concept analysis, ensuring that a
sufficient number of mocks satisfy
\(S_{{\rm best}}\geq S_{\rm best}^{\rm obs}\).  A real-data analysis may
require an even larger mock ensemble to obtain a more precise \(p\)-value.
The
\(\mathcal T=2\) chain is then reweighted for each mock using
Eq.~\eqref{eq:mock-reweight}, and the resulting \(S_{{\rm best},b}\)
ensemble is used to evaluate Eq.~\eqref{eq:ppc-pvalue}.  The corresponding
posterior-predictive calibration results are summarized in
Table~\ref{tab:posterior-results}.

\begin{table}[t]
\centering
\begin{tabular}{lccccc}
\toprule
Case & truth & \(N_{\rm mock}\) & \(S_{\rm best}^{\rm obs}\) & \(p_{\rm PPC}\) & \(Z_{\rm 1s}\) \\
\midrule
Null & \(A(k)=1\) & 1000 & \(6.31\) & \(0.404\) & \(0.24\sigma\) \\
1\% modulation & Eq.~\eqref{eq:oscillatory-modulation} with \(\mathcal A_{\rm osc}=0.01\) & 10000 & \(24.85\) & \(0.0028\) & \(2.77\sigma\) \\
\bottomrule
\end{tabular}
\caption{Posterior-weighted consistency-test quantities for the null and
1\% modulation cases.  The 1\% row uses
Eq.~\eqref{eq:oscillatory-modulation} with
\(\mathcal A_{\rm osc}=0.01\).  Both rows use the same best-\(\chi^2\)
RegMRL rule and mock-posterior reweighting.  The final column gives the one-sided
upper-tail Gaussian equivalent defined in Eq.~\eqref{eq:one-sided-sigma}.}
\label{tab:posterior-results}
\end{table}

For the null case, \(S_{\rm best}^{\rm obs}=6.31\).  In the corresponding
posterior-predictive ensemble, 403 out of 1000 null mocks have
\(S_{\rm best}\ge S_{\rm best}^{\rm obs}\), giving the plus-one estimate
\(p_{\rm PPC}=0.404\), or \(Z_{\rm 1s}=0.24\sigma\).  The
likelihood-ratio effective sample size for the observed statistic is
\(N_{\rm eff}=113.85\), indicating that the
observed statistics are not dominated by only a few posterior samples

\begin{figure}[t]
\centering
\IfFileExists{figures/posterior_weighted/S_best_distribution_vs_direct_null_nodrag.pdf}{%
  \includegraphics[width=0.78\textwidth]{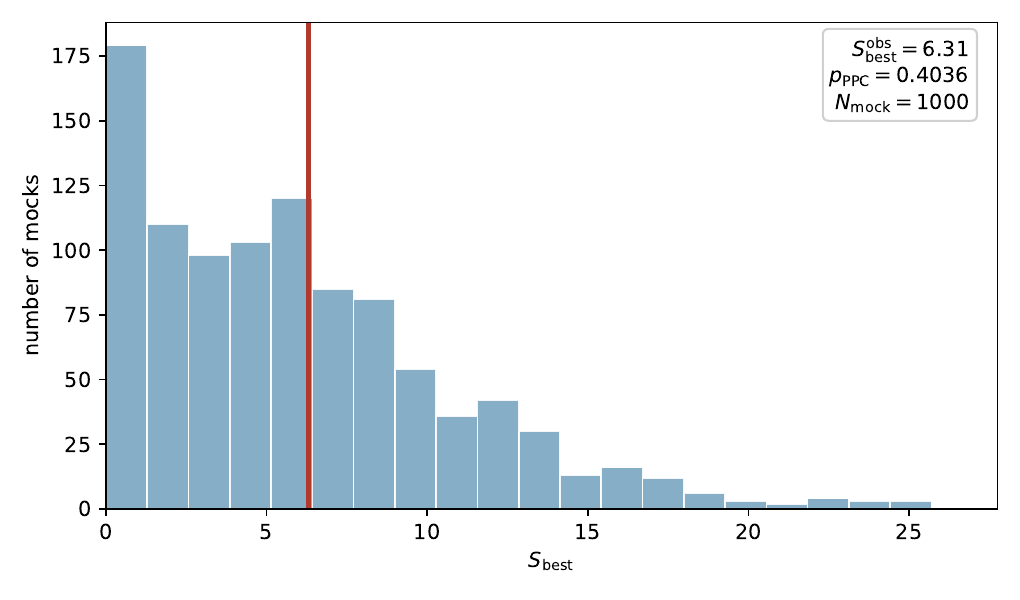}%
}{%
  \fbox{\parbox[c][0.34\textheight][c]{0.78\textwidth}{\centering
  Placeholder for the \(S_{\rm best}\) posterior-predictive distribution in
  the null case, with \(S_{\rm best}^{\rm obs}\) marked.}}%
}
\caption{Posterior-weighted \(S_{\rm best}\) calibration for the null case.
The vertical red line marks \(S_{\rm best}^{\rm obs}=6.31\).  The plus-one
tail probability is \(p_{\rm PPC}=0.404\), corresponding to
\(Z_{\rm 1s}=0.24\sigma\).}
\label{fig:sbest-null-planned}
\end{figure}

For the 1\% modulation case, evaluating the observed statistic gives
\(S_{\rm best}^{\rm obs}=24.85\).  In the posterior-predictive null ensemble,
27 out of 10000 mocks have \(S_{\rm best}\ge S_{\rm best}^{\rm obs}\), giving
\(p_{\rm PPC}=0.0028\), or \(Z_{\rm 1s}=2.77\sigma\).  The likelihood-ratio
effective sample size for the observed statistic is
\(N_{\rm eff}=948.93\).

\begin{figure}[t]
\centering
\includegraphics[width=0.78\textwidth]{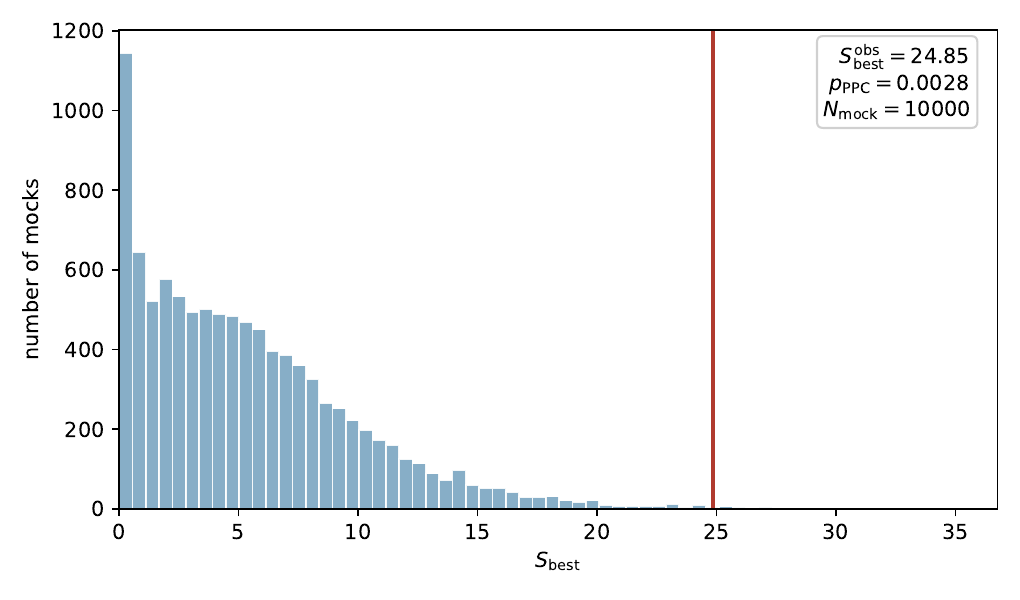}
\caption{Posterior-predictive \(S_{\rm best}\) null calibration ensemble for
the 1\% modulation case.  The vertical red line marks the observed statistic, \(S_{\rm best}^{\rm obs}=24.85\), with
\(p_{\rm PPC}=0.0028\), corresponding to \(Z_{\rm 1s}=2.77\sigma\).}
\label{fig:sbest-onepercent}
\end{figure}

The null realization result shows that the
RegMRL improvement expected from noise and posterior uncertainty is not being
mistaken for evidence against \(A(k)=1\).  By contrast, the 1\% oscillatory
case lies in the upper tail of the same null-calibrated statistic, indicating
that this Rubin/LSST-Y10-like configuration is sensitive to a coherent percent-level
modulation of the assumed shape.

\section{Conclusion}
\label{sec:conclusion}

We have developed a deconvolution-based framework for testing
scale-dependent freedom in the late-time matter power spectrum with
3x2pt measurements.  The matter spectrum is written as a
scale-only modulation \(A(k)\) of a fiducial nonlinear spectrum, and the
full-sky \(gg\), \(g\gamma\), and \(\gamma\gamma\) kernels are assembled into
a linear response matrix for binned band powers.  This gives a
forward model in which density, redshift-space distortion, gravitational
shear, and intrinsic-alignment terms all contribute to the same reconstructed
function \(A(k)\).

We adapted the regularized modified Richardson--Lucy update to
the 3x2pt response matrix with a weak diffusion term.  The
reported reconstruction is selected by the minimum $\chi^2$ along the
iteration history.  For the Rubin/LSST Y10-like configuration considered here, a \(5\%\)
oscillatory modulation can be reliably recovered from noisy realizations over
\(10^{-1}\,{\rm Mpc}^{-1}\lesssim k\lesssim5\times10^{-1}\,{\rm Mpc}^{-1}\).
On larger
physical scales the ensemble median can still follow the true \(A(k)\), but the
large realization-to-realization scatter makes the reconstruction unreliable
for a single observed data set.  For a \(1\%\) feature, the noisy realizations
show only a collective departure from \(A(k)=1\) over approximately the same
range, with several realizations remaining close to the null curve.  This
identifies the \(1\%\) level as the approximate sensitivity of the adopted
observational configuration for this assumed oscillatory shape.

We also introduce a posterior-weighted consistency statistic \(S_{\rm best}\).  We evaluate the
observed \(S_{\rm best}\) on the posterior, generate noisy
posterior-predictive mock data from the same chain, and apply the same
best-\(\chi^2\) RegMRL rule to each mock.  This calibration avoids assigning a Wilks
\(p\)-value to a statistic whose effective dimensionality is determined by the
regularization, clipping, and iteration selection.

For the null synthetic data, the observed statistic is typical of its
posterior-predictive null ensemble, corresponding to a one-sided Gaussian
equivalent of \(0.24\sigma\).  For the \(1\%\) oscillatory modulation, the same
calibration procedure gives a one-sided Gaussian equivalent of
\(2.77\sigma\).  These results show that the method can distinguish a coherent
weak modulation from the look-elsewhere behavior of the regularized
deconvolution in controlled Rubin/LSST-Y10-like synthetic data.

The present analysis is a proof of concept rather than a claim about
real survey data.  Applications to real data would rely on collaboration-validated data products and systematic-uncertainty treatments, including those associated with photometric redshifts and shear calibration, while our modeling would need to account more fully for baryonic feedback, intrinsic alignments, and possible redshift dependence in
\(A(k,z)\).  Finally, combining the 3x2pt kernels studied here with
the CMB-lensing reconstruction developed in the first paper of this series
would turn the method into a joint low- and high-redshift test of coherent
matter-power-spectrum features.

\acknowledgments
We thank Wuhyun Sohn for meaningful discussions.
J.J. and A.S. acknowledge funding from the Korea Astronomy and Space Science Institute (KASI) through the project “Research on the Principles of the Accelerating Expansion of the Universe” (Project Code: 2026183201).
D.K.H. acknowledges financial support from the Indo-French Centre for the Promotion of Advanced Research (IFCPAR/CEFIPRA), New Delhi, India, through the Collaborative Scientific Research Programme (Project No. 6704-4, “Testing flavors of the early universe beyond vanilla models with cosmological observations”); and from the Anusandhan National Research Foundation (ANRF), Government of India, under the ARG MATRICS scheme (Grant No. ANRF/ARGM/2025/000941/TS; project “EPOCH: Exploring Primordial Origins in Cosmic Hierarchies”).

\appendix

\section{Survey configuration}
\label{app:survey-config}

We adopt a Rubin/LSST Year 10 (Y10)-like harmonic-space 3x2pt configuration,
following the broad survey and systematic-error assumptions of the Rubin/LSST DESC
Science Requirements Document (SRD)~\cite{LSSTDESC:2018kqk}.  The source
sample is split into five tomographic bins with total effective number density
\(27\,{\rm arcmin}^{-2}\) and per-component shape noise
\(\sigma_e=0.26\).  The lens sample is split into ten tomographic bins with
total number density \(48\,{\rm arcmin}^{-2}\).  Both source and lens
redshift distributions are taken from the Y10 SRD tabulations.  The covariance
is the SRD Y10 3x2pt covariance for a survey area of
\(14202.63\,{\rm deg}^2\), and we use top-hat bandpower windows.

The data vector contains source-shear auto- and cross-spectra,
galaxy--galaxy lensing spectra, and lens-galaxy clustering spectra.  For
cosmic shear we use all \(5(5+1)/2=15\) source-bin pairs.  For
galaxy--galaxy lensing we keep the lens--source pairs for which the source bin
lies behind the lens bin in the SRD binning, giving 25 tracer pairs.  For
galaxy clustering we use the ten lens-bin auto-spectra.  All three statistics
are binned on the same logarithmic multipole grid beginning at
\(\ell=20\); the density-containing statistics are additionally subject to
the scale cut \(k_{\max}=0.201\,{\rm Mpc}^{-1}\), while cosmic
shear has no additional \(k_{\max}\) cut.  After applying these cuts the
synthetic data vector has 557 retained band powers.

The tracer definitions, redshift distributions, covariance, scale cuts, and
bandpower windows are generated with Augur and then evaluated with the
full-sky forward model described above.  The number-count tracer in lens bin
\(i\) contains a density term with linear galaxy bias \(b_i\), a
photometric-redshift shift \(\Delta z_{{\rm lens},i}\), and the RSD term in
Eq.~\eqref{eq:phi-rsd}.  The source-shape tracer in source bin \(i\) contains
gravitational lensing, a photometric-redshift shift
\(\Delta z_{{\rm src},i}\), a multiplicative shear calibration \(m_i\), and
an intrinsic-alignment contribution.  CMB-lensing tracers are not included in
the baseline analysis.  For the direct posterior fits used in the consistency
test, the same survey definition is combined with a CAMB~\cite{Lewis:1999bs} background and
matter-power calculation, using the Mead2020~\cite{Mead:2020vgs} nonlinear prescription and one
fixed massive-neutrino species with total mass \(0.06\,{\rm eV}\).

The intrinsic-alignment contribution is modeled with a nonlinear-alignment
(NLA) form~\cite{Catelan:2000vm,Hirata:2004gc,Bridle:2007ft}.  We write the IA transfer amplitude entering
Eq.~\eqref{eq:phi-ia} as
\begin{equation}
    F_{\rm IA}(z)
    =
    -\,A_{\rm IA}^{\rm eff}(z)\,
    C_1\rho_{\rm crit}\Omega_m\,D_{\rm lin}^{-1}(z),
    \label{eq:nla-fia}
\end{equation}
where \(D_{\rm lin}(z)\) is the scale-independent linear growth factor
normalized to unity today, \(C_1=5\times10^{-14}\,
h^{-2}M_\odot^{-1}{\rm Mpc}^3\), and \(\rho_{\rm crit}\) is the critical
density today.  The redshift-dependent dimensionless amplitude is
\begin{align}
    A_{\rm IA}^{\rm eff}(z)
    &=
    A_{\rm IA}
    \left(\frac{1+z}{1+z_{\rm piv}}\right)^{\alpha_{\rm IA}}
    D_{\rm lin}^{\alpha_g-1}(z)\,
    H_{\rm IA}(z),\notag\\
    H_{\rm IA}(z)
    &=
    \Theta(z_{\rm br}-z)
    +
    \Theta(z-z_{\rm br})
    \left(\frac{1+z}{1+z_{\rm br}}\right)^{\eta_{\rm high-z}},
    \label{eq:nla-aia}
\end{align}
with \(z_{\rm piv}=0.3\), \(z_{\rm br}=0.7\), and \(\alpha_g=1\) in the
baseline configuration.  The IA nuisance parameters are therefore the
amplitude \(A_{\rm IA}\), the redshift-scaling exponent \(\alpha_{\rm IA}\),
and the high-redshift modifier \(\eta_{\rm high-z}\).

The cosmological posterior used for fit varies
\(\log(10^{10}A_s)\), \(\Omega_m\), \(H_0\), \(n_s\), and
\(\omega_b\equiv\Omega_bh^2\), with \(A_s\) sampled through
\(\log(10^{10}A_s)\).  We use broad top-hat priors on
\(\log(10^{10}A_s)\), \(\Omega_m\), \(H_0\), and \(n_s\), and a Gaussian
prior on \(\omega_b\) centered on the fiducial value\footnote{The prior on $\omega_b$ is inspired by the current BBN constraint on the baryon abundance, e.g.~\cite{Schoneberg:2024ifp}}:
\begin{align}
    \log(10^{10}A_s)&\sim \mathcal U(1.61,3.91),&
    \Omega_m&\sim \mathcal U(0.1,0.5),\notag\\
    H_0&\sim \mathcal U(30,100)\ {\rm km\,s^{-1}\,Mpc^{-1}},&
    n_s&\sim \mathcal U(0.8,1.2),\notag\\
    \omega_b&\sim \mathcal N(\omega_{b,{\rm fid}},\,0.00055^2).
    \label{eq:survey-cosmo-priors}
\end{align}
The helium fraction, CMB temperature, effective number of relativistic
species, and neutrino mass are fixed in these fits.  The cold-dark-matter
density is derived from \(\Omega_m\), \(\Omega_b\), and \(h\), so that the
sampled model remains in the flat-\(\LCDM\) family.

The nuisance-parameter priors are the Gaussian priors used in the
3x2pt likelihood:
\begin{align}
    A_{\rm IA}&\sim \mathcal N(1,5^2),&
    \alpha_{\rm IA}&\sim \mathcal N(0,2.3^2),&
    \eta_{\rm high-z}&\sim \mathcal N(0,0.8^2),\notag\\
    b_i&\sim \mathcal N(b_{i,{\rm fid}},0.9^2),&
    \Delta z_{{\rm src},i}&\sim \mathcal N(0,0.002^2),&
    \Delta z_{{\rm lens},i}&\sim \mathcal N(0,0.005^2).
    \label{eq:survey-nuisance-priors}
\end{align}
Here \(i=1,\ldots,10\) for the lens-bin galaxy biases and lens redshift
shifts, while \(i=1,\ldots,5\) for the source redshift shifts.  The parameters
\(A_{\rm IA}\), \(\alpha_{\rm IA}\), and \(\eta_{\rm high-z}\) are the IA
parameters defined in Eqs.~\eqref{eq:nla-fia}--\eqref{eq:nla-aia}.  The
fiducial linear galaxy-bias centers are
\begin{align}
    (b_{i,{\rm fid}})_{i=1}^{10}
    =
    &(1.3767,1.4512,1.5284,1.6080,1.6896,\notag\\
    &1.7729,1.8577,1.9438,2.0309,2.1189).
    \label{eq:lens-bias-fiducials}
\end{align}
The fiducial photometric-redshift shifts and shear-calibration biases are
zero.

For source multiplicative shear calibration we follow the redshift-dependent template rather than sampling five independent \(m_i\).  A single
amplitude \(m_0\) is assigned the prior
\begin{equation}
    m_0\sim \mathcal N(0,0.013^2),
\end{equation}

and the source-bin multiplicative biases are
\begin{equation}
    m_i
    =
    m_0\,
    \frac{2\bar z_i-\bar z_{\rm max}}{\bar z_{\rm max}},
    \qquad
    \bar z_i=(0.309,0.589,0.867,1.241,2.053),
    \qquad
    \bar z_{\rm max}=2.053.
    \label{eq:srd-multiplicative-bias-template}
\end{equation}
This makes the shear-calibration uncertainty coherent across source bins while
allowing its sign and magnitude to vary with source redshift.

The synthetic data vectors are generated at the fiducial parameter values,
which coincide with the centers of the systematics priors.  The cosmological
fiducial is a Planck18-like flat-\(\LCDM\) model,
\begin{align}
    \Omega_c&=0.264470,\quad
    \Omega_b=0.049302,\quad
    h=0.6736,\quad
    n_s=0.9649,\notag\\
    A_s&=2.0989\times10^{-9},\quad
    \sum m_\nu=0.06\,{\rm eV},\quad
    N_{\rm eff}=3.046 .
    \label{eq:synthetic-fiducial-cosmology}
\end{align}
The systematics are set to their prior centers:
\(\Delta z_{{\rm src},i}=\Delta z_{{\rm lens},i}=0\),
\(m_0=0\), \(m_i=0\), \(A_{\rm IA}=1\), \(\alpha_{\rm IA}=0\), and
\(\eta_{\rm high-z}=0\), with the lens biases fixed to
Eq.~\eqref{eq:lens-bias-fiducials}.

\section{Modulation frequency and \texorpdfstring{\(k\)}{k}-sampling checks}
\label{app:kgrid-frequency-checks}

In this appendix, we repeat the noisy-ensemble
reconstruction for additional injected shapes.  We consider logarithmic-phase
oscillations,
\begin{equation}
    A_{\log}(k;f)
    =
    1+\mathcal A_{\rm osc}
    \sin\!\left[
    f\log_{10}\left(\frac{k}{k_\star}\right)
    \right],
\end{equation}
and a linear-\(k\) phase,
\begin{equation}
    A_{\rm lin}(k;f)
    =
    1+\mathcal A_{\rm osc}
    \sin\!\left[
    f\left(\frac{k}{k_\star}-1\right)
    \right],
\end{equation}
with \(k_\star=0.2\,{\rm Mpc}^{-1}\) and
\(\mathcal A_{\rm osc}=0.05\).

\begin{figure}[H]
\centering
\includegraphics[width=0.82\textwidth]{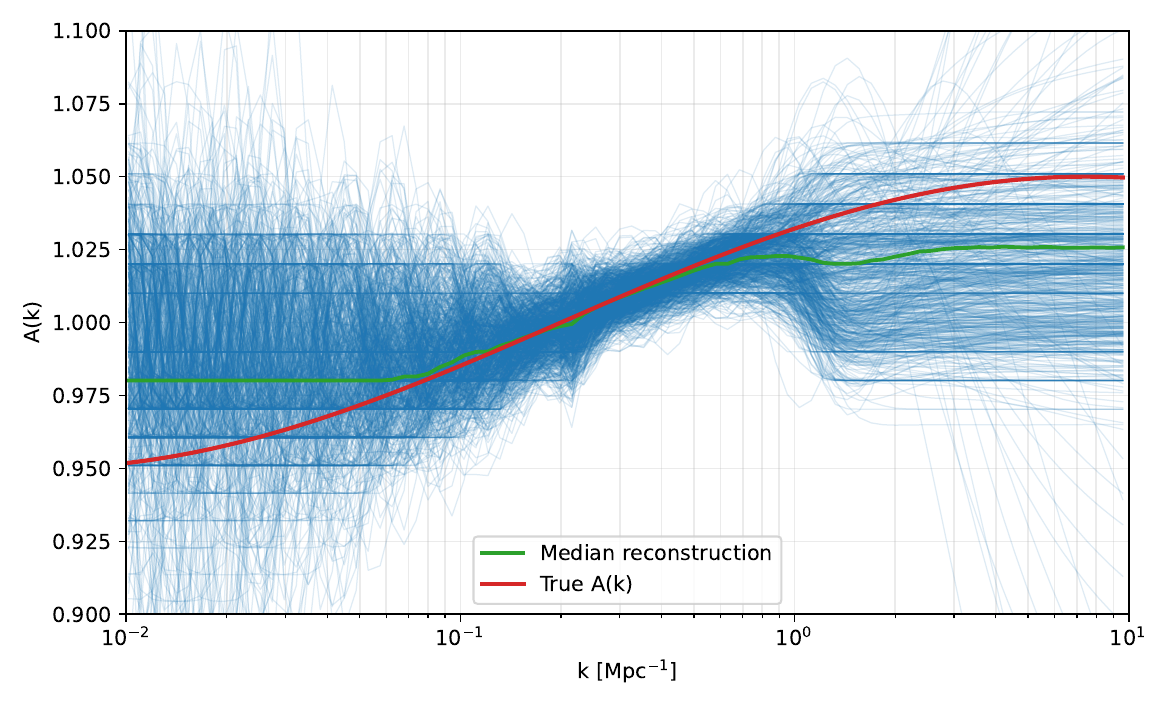}
\caption{RegMRL reconstruction ensemble for a 5\% logarithmic-phase
modulation with \(f=1\), using the baseline \(N_k=160\) grid and
\(\kappa=10^{-3}\).}
\label{fig:appendix-logfreq1}
\end{figure}

\begin{figure}[H]
\centering
\includegraphics[width=0.82\textwidth]{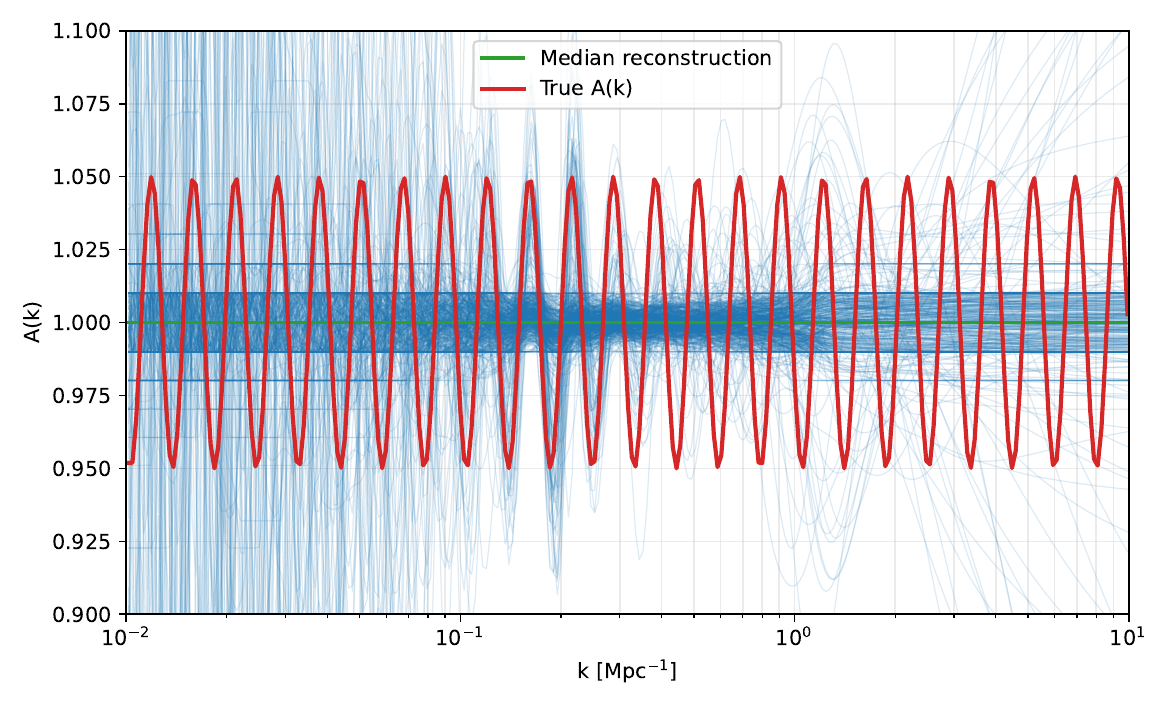}
\caption{RegMRL reconstruction ensemble for a 5\% logarithmic-phase
modulation with \(f=50\).  We use \(N_k=512\) and
\(\kappa=10^{-5}\) so that the injected feature is resolved by the grid and is
not removed by the explicit diffusion step.}
\label{fig:appendix-logfreq50}
\end{figure}

\begin{figure}[H]
\centering
\includegraphics[width=0.82\textwidth]{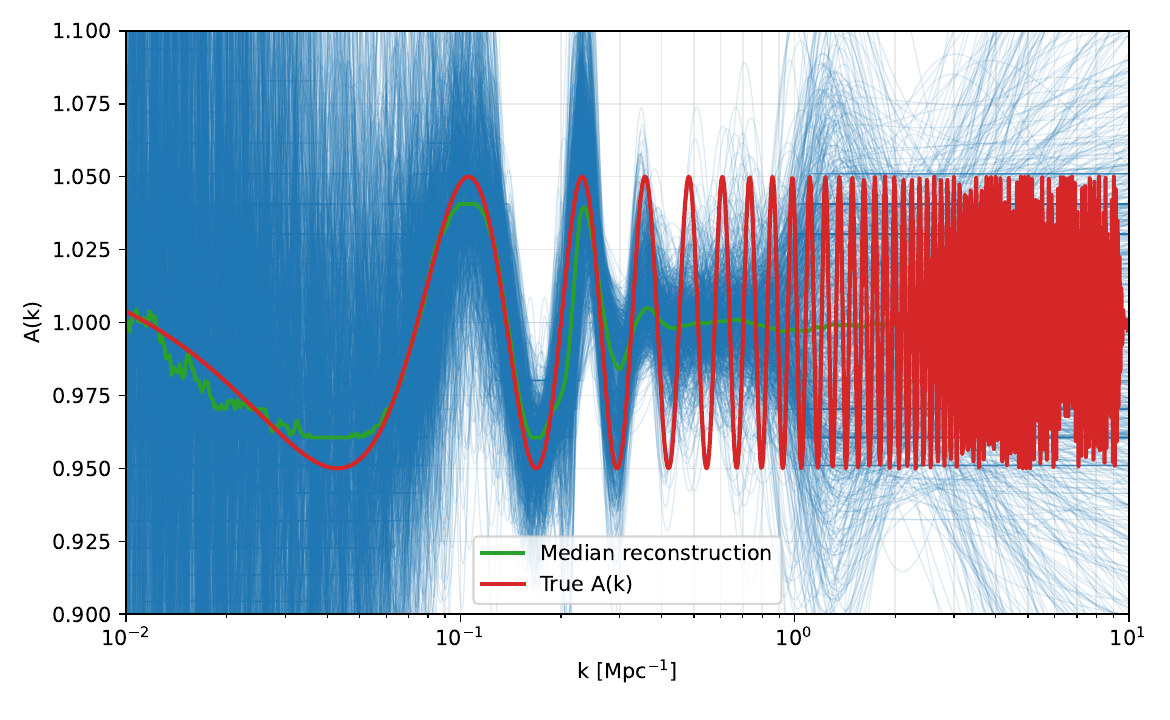}
\caption{RegMRL reconstruction ensemble for a 5\% linear-\(k\)
phase modulation with \(f=10\), using \(N_k=2048\) and
\(\kappa=10^{-5}\).}
\label{fig:appendix-lineark}
\end{figure}

Figure~\ref{fig:appendix-logfreq1} shows that a slowly varying logarithmic
modulation with $f=1$ can be recovered over the same best-constrained range
identified in the main text.
Within the sensitive scale range
\(10^{-1}\,{\rm Mpc}^{-1}\lesssim k\lesssim
5\times10^{-1}\,{\rm Mpc}^{-1}\), the modulation has an effective amplitude
of only~$\mathcal{O}(0.01)$ and therefore lies close to the threshold for
distinguishing it from $A(k)=1$.
Figure~\ref{fig:appendix-logfreq50} shows the high-frequency logarithmic case
with \(f=50\).  For this case, we refine the grid to \(N_k=512\), ensuring that
the sampling rate safely exceeds the Nyquist requirement, and reduce the
regularization strength to \(\kappa=10^{-5}\), making the effective diffusion
scale smaller than the feature scale.
Although a few noisy realizations recover the oscillatory pattern near
$k\sim0.2\,{\rm Mpc}^{-1}$, the ensemble median shows that the feature is not
recovered in most cases because it is smoothed by the lensing convolution
kernels.  Figure~\ref{fig:appendix-lineark} shows the result for a linear-$k$
modulation.  We increase the grid resolution to \(N_k=2048\) and use
\(\kappa=10^{-5}\).
Since the oscillation frequency near the pivot scale
$k_\star\sim0.2\,{\rm Mpc}^{-1}$ is comparable to that in the baseline case
and decreases toward larger scales, the feature can be recovered over
\(10^{-1}\,{\rm Mpc}^{-1}\lesssim k\lesssim
2\times10^{-1}\,{\rm Mpc}^{-1}\).  These tests indicate that, in the interval
\(10^{-1}\,{\rm Mpc}^{-1}\lesssim k\lesssim
5\times10^{-1}\,{\rm Mpc}^{-1}\), the reconstruction is effective for
oscillatory features with frequency \(f\lesssim 10\) on \(\log_{10}(k/0.2\,{\rm Mpc}^{-1})\).  More rapidly varying features
are smoothed by the convolution kernels before the deconvolution step can
recover them.

\bibliographystyle{JHEP}
\bibliography{biblio}

\end{document}